\documentclass[a4paper,fleqn,usenatbib]{mnras}
\usepackage{float}
\usepackage[T1]{fontenc}
\usepackage{ae,aecompl}
\usepackage{graphicx}
\usepackage{amsmath}
\usepackage{amssymb}
\usepackage{supertabular}
\usepackage{siunitx}
\usepackage{caption}
\DeclareCaptionLabelFormat{cont}{Figure#1~#2\alph{ContinuedFloat}}
\def\beq{\begin{eqnarray}}
\def\eeq{\end{eqnarray}}

\title[Neutrinos and GWs from NDAFs in fallback CCSNe]
{Anisotropic neutrinos and gravitational waves from black hole neutrino-dominated accretion flows in fallback core-collapse supernovae}

\author[Wei, Liu $\&$ Xue]
{Yun-Feng Wei, Tong Liu\thanks{E-mail: tongliu@xmu.edu.cn}, Li Xue \\
Department of Astronomy, Xiamen University, Xiamen, Fujian 361005, China}

\date{Accepted XXX. Received YYY; in original form ZZZ}

\pubyear{2021}

\begin{document}

\maketitle

\begin{abstract}
Fallback in core-collapse supernovae (CCSNe) plays an important role in determining the properties of the central compact remnants, which might produce a black hole (BH) hyperaccretion system in the centre of a massive CCSN. When the accretion rate is extremely high and neutrino cooling is dominant, the hyperaccretion should be in the phase of the neutrino-dominated accretion flows (NDAFs), and thus a large number of anisotropic MeV neutrinos will be launched from the disc along with the strong gravitational waves (GWs). In this paper, we perform a series of one-dimensional CCSN simulations with the initial explosion energy in the range of $2-8$ B (1 B = $10^{51}$ erg) to investigate the fallback processes. By considering the evolution of the central BH mass and spin in the fallback accretion, we present the effects of the initial explosion energies, masses and metallicities of the massive progenitor stars on the spectra of anisotropic MeV neutrinos and the waveform of GWs from NDAFs. These neutrino or GW signals might be detected by operational or future detectors, and the multimessenger joint detections could constrain the properties of CCSNe and progenitor stars.
\end{abstract}


\section{Introduction}\label{sec:intro}

Massive stars ($> 8~ M_{\odot}$) usually end their lives as a core-collapse supernova (CCSN). The explosions give birth to neutron stars (NSs) or black holes (BHs) and eject solar masses of heavy elements \citep[see e.g.][]{Colgate1966,Burrows2021}. Although the CCSN theory has been studied for more than half a century and has achieved remarkable progress, the explosion mechanism remains uncertain. The delayed neutrino-heating mechanism is considered a robust solution and has been widely explored since its conception by \citet{Colgate1971}. The collapse of an iron core in the centre of a massive star ($\sim 20-40 ~M_\odot$) initially leads to the product of a proto-NS and the formation of a shock wave after the core reaches the nuclear density and rebounds. The shock wave travels outwards and loses energy by dissociating iron nuclei and stalls. The CCSN then enters the accretion phase, in which infalling matter accretes to the shock front and the proto-NS continues to grow \citep[e.g.][]{Bethe1990,Walk2020}. In the paradigm of neutrino-driven explosions, the shock is then revived by the joint action of neutrino heating and various hydrodynamic instabilities and a successful CCSN appears \citep[e.g.][]{Bethe1985,Mueller2016,Janka2007,Janka2012}. A successful explosion, in which the stellar mantle is ejected, will result in the formation of an NS. In many cases, however, the remnant will be a BH. A considerable amount of material (the material that does not reach escape velocities, or that is decelerated by a subsequent reverse shock) falls back, causing the NS to collapse to a BH \citep[e.g.][]{Chevalier1989,Fryer2006,Li2020}. CCSN fallback plays an important role in determining the properties of the compact remnants and of the ejecta composition \citep[e.g.][]{Chan2018}. Meanwhile, fallback might be related to a number of the observed phenomena, i.e., the peculiar supernovae (SNe), the late-time neutrino emission, the $r$-process element productions, and the long-duration gamma-ray bursts \citep[LGRBs, e.g.][]{Wong2014,Liu2018}.

The concept of fallback was first discussed by \citet{Colgate1971}. Since that time, many CCSN explosion calculations have confirmed the existence of fallback and studied its dynamics and effects \citep[see e.g.][]{Bisnovatyi1984,Woosley1989,Chevalier1989,Woosley1995,Fryer1999,Fryer2009,MacFadyen2001,Zhang2008,Moriya2010,Moriya2019,Dexter2013,Wong2014,Perna2014,Branch2017,Chan2020}. Moreover, substantial observational evidences for CCSN fallback had been gathered \citep[e.g.][]{Israelian1999,Zampieri2003,Moriya2010,Moriya2018,Nomoto2006,Keller2014,Bessell2015}. The intensity of the fallback is determined by the CCSN explosion energy and the binding energy of the star \citep[e.g.][]{Fryer2006}. The more powerful fallback may correspond to the weaker explosion energy. Therefore, the electromagnetic signals from the fallback CCSNe would be faint and might be undetectable. Nevertheless, neutrinos and gravitational waves (GWs) can be the unique probes of the core-collapse of massive stars. They can provide useful information regarding the fallback processes.

Neutrinos from fallback in CCSNe have been studied in \citet{Fryer2009}. He calculated neutrinos from the fallback onto the newly formed NS and showed that the fallback can contribute a sizeable fraction of the total observed neutrino flux. For the rotating progenitors, the fallback could provide power to the central engine producing LGRBs \citep[e.g.][]{Woosley1993,MacFadyen1999,Woosley2012,Liu2018,Liu2019}. Infalling material with enough angular momentum would be slowed by the rotation and be piled into a disc. The fallback accretion rate is high and the disc would be a hyperaccretion disc once the temperature and density are high enough that the photons are trapped and large amounts of neutrinos are emitted. Such a disc would be in a state of neutrino-dominated accretion flow \citep[NDAF, see e.g.][]{Popham1999,Narayan2001,Kohri2002,Lee2005,Gu2006,Chen2007,Janiuk2007,Kawanaka2007,Liu2007,Liu2015,Liu2016,Lei2009,Li2013,Luo2013,Xue2013,Song2016,Nagataki2018}, and for review see \citet{Liu2017a}. The anisotropic neutrino emission from the disc would then generate the GW emission \citep[e.g.][]{Suwa2009,Liu2017b,Wei2020}.

The explosion can be parameterized by the motion of a piston or by injecting a prescribed amount of energy into the inner zone. Both mechanisms have been used to investigate the fallback processes \citep[e.g.][]{Zhang2008,Fryer2009,Dexter2013}. The fallback from the piston-driven explosion is different to that from the energy-driven explosion \citep[e.g.][]{Young2007,Wong2014}. The piston engine is still a useful tool to investigate the fallback CCSNe in some recent works \citep[e.g.][]{Sukhbold2016,Woosley2019}.

In this paper, we adopt the piston engine to simulate the fallback processes. We then roughly use fallback rate to estimate the mass accretion rate of the centre hyperaccretion disc in order to investigate the anisotropic neutrino and GW radiations. This paper is organized as follows. In Section 2, we present the CCSN simulation methods and results. The evolution of a BH in a hyperaccretion system is introduced in Section 3. In Sections 4 and 5, we calculate the neutrino and GW spectra of NDAFs in the centre of CCSNe, respectively. The effects of the initial explosion energies, the viewing angles and the masses and metallicities of the progenitor stars are studied. A brief summary is given in Section 6.

\section{CCSN simulations}

\begin{figure*}
\centering
\includegraphics[width=15cm,height=10cm]{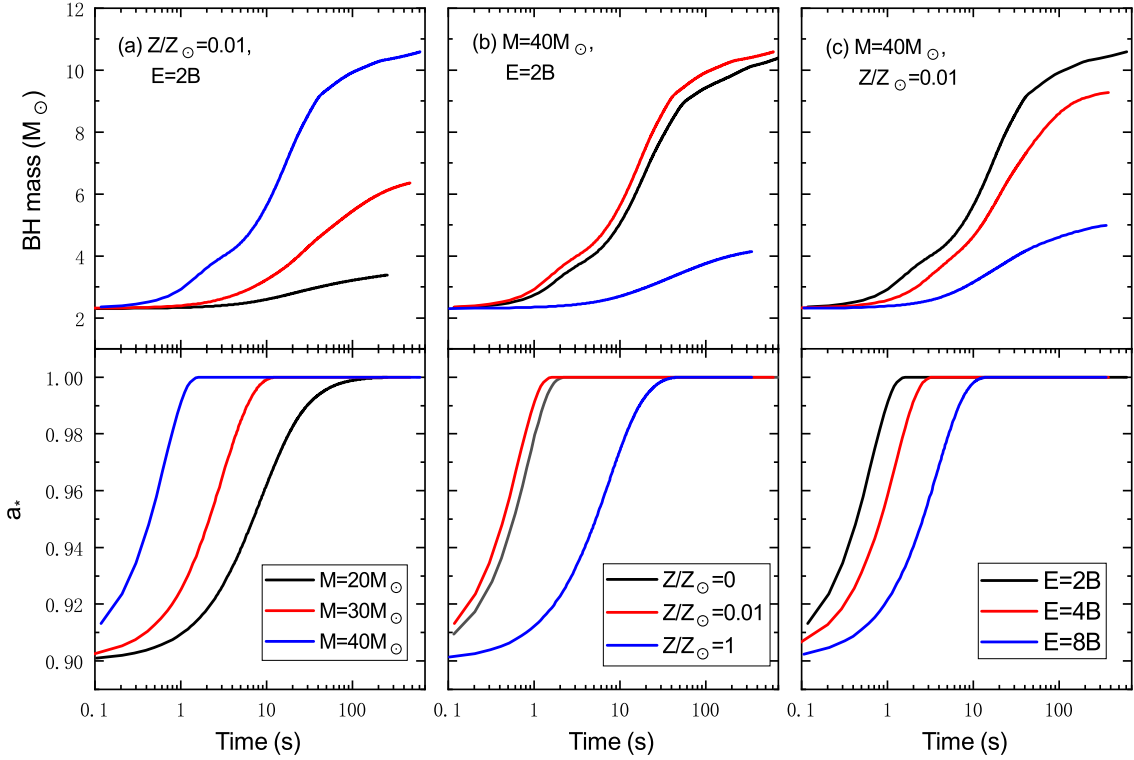}
\caption{Evolutions of mass and spin of BHs surrounded by hyperaccretion discs with different progenitor masses, metallicities, and initial explosion energies. The initial BH mass and spin are set as $M_{\rm{BH}}=2.3 ~M_{\odot}$ and $a_{*}=0.9$, respectively.}
\label{figure1}
\end{figure*}

In this work, we adopt the well-known pre-SN models \citep[e.g.][]{Woosley2002,Woosley2007,Heger2010} as the initial conditions. These progenitor models with initial mass in range of $20-40 M_{\odot}$ were evolved using the KEPLER code \citep{Weaver1978,Woosley2002} through all stable stages of nuclear burning until their iron cores became unstable and collapsible. Amounts of these models, the ones with zero-metallicity ($Z/Z_\odot=0$) and solar-metallicity ($Z/Z_\odot=1$) are referenced from \citet{Heger2010} and \citet{Woosley2007}, respectively, as well as the ones with metallicity $Z/Z_\odot=0.01$ are provided by Prof. Alexander Heger in private communication. Here $Z$ and $Z_{\odot}$ are the metallicities of the progenitor stars and the Sun, respectively. We adopt the Athena++ code \citep{White2016} to perform one-dimensional CCSN simulations with the fixed inner boundary at $R$ = $10^{9}$ cm. The initial explosion energies are respectively set to be $E$ = 2, 4, and 8 B (1 B = $10^{51}$ erg) for different cases. In order to simplify the initial explosion within the inner boundary, which is out of the scope of this paper, we follow \citet{Woosley1995} and \citet{Woosley2002} to adopt the piston approach to mimic it. In all of our simulations, the piston is initially located at the outer edge of the iron core, and the piston firstly moves inwards for 0.45 s when the collapse begins, and then abruptly moves outwards at a certain small radius with an initial high velocity and decelerates smoothly until coming to rest at $10^{9}$ cm.

For each case, the simulation was divided into two steps to model the initial collapse and the subsequent explosion accompanied by the fallback process. In the first step, the structural profiles of progenitor stars were mapped into the Athena++ code. The outer boundary of the computational domain is set at the surface of the progenitor star, which is different for each case (from approximately $10^{12}$ to $10^{14}$ cm). A unidirectional outflowing inner boundary condition was used to mimic the suction effect resulting from the hypothetical piston moving inwards. This step briefly reflects a free collapse of the star before the beginning of the explosion. The simulation is run to 0.45 s, after which the piston turns outwards, corresponding to the outward propagation of the blast. An average of approximately 1 s is required for a blast wave to reach $R$ = $10^{9}$ cm \citep{Burrows2020}, which is also the period in which the piston moves outwards. The gas flow does not change much during this brief period \citep{Liu2021}. Thus, we directly map the star shape at 0.45 s to the new grid for the second step as its initial conditions.

In the second step, the same outflowing inner boundary is set at $R$ = $10^{9}$ cm, while the outer boundary is set at $R$ = $10^{16}$ cm, which is far from the star surface. Outside the star, the medium is maintained in a constant state with pressure and density three orders of magnitude lower than the corresponding values on the star surface. In order to mimic the outward blast passing through the inner boundary, the additional energy and mass are artificially injected into the innermost computing cell at the beginning of the second step. This injection is assumed to be instantaneous, so it only needs to modify the initial condition of the inner most cell rather than setting a time-dependent boundary condition. The injected energy is just the setting explosion energy. There are only three values taken in our simulations, namely 2, 4, and 8 B. The injected mass consists of two fractions. One fraction is from the recording of inhaled mass during 0.45 s collapse in the first step. The other fraction is the mass within the inner boundary ($\sim 10^{9}$ cm) excluding the mass of the iron core. For different progenitor stars, the injected mass value is in the range of $\sim 0.5-2.5~M_{\odot}$. All of simulations in the second step were run until the remnant growth ceased and the maximum duration was about $3\times 10^{6}$ s.

For the two different steps above, the grid for each step has a different number of cells. In the first step, the grid has $10^{4}$ cells with a logarithmic uniform interval for the radial direction. To reduce computing time, the grid for the second step has only 2,000 logarithmic spacing cells.

The profiles of the density and velocity at 50, 100, 500, and 1,000 s and the time evolutions of the mass supply rate for different initial explosion energies with the different progenitor masses and metallicities are shown in Appendix A. The effects of different initial conditions on explosion evolution have been discussed in \citet{Liu2021}. In that work, we mainly focused on the properties of final compact remnants of CCSNe and investigated the existence of the lower mass gap in the compact object distribution. Here, we mainly focus on the evolution of the fallback mass supply rates. As shown in Figure A3, for the same explosion energy, more powerful fallback would occur in the collapse of the progenitor star with higher mass and lower metallicity. For a given progenitor star, the mass supply rate decreases significantly as the initial explosion energy increases. Without considering the disc outflows, the mass supply rate can be roughly considered as the net accretion rate. We notice that in all cases, the mass accretion rates are approximately in the range of $0.01-1~M_{\odot }\rm ~s^{-1}$ during the initial stage. With such a high mass accretion rate, the hyperaccretion disc would be in the state of the NDAF.

\section{BH evolution}

\begin{figure*}
\centering
\includegraphics[width=8cm,height=6cm]{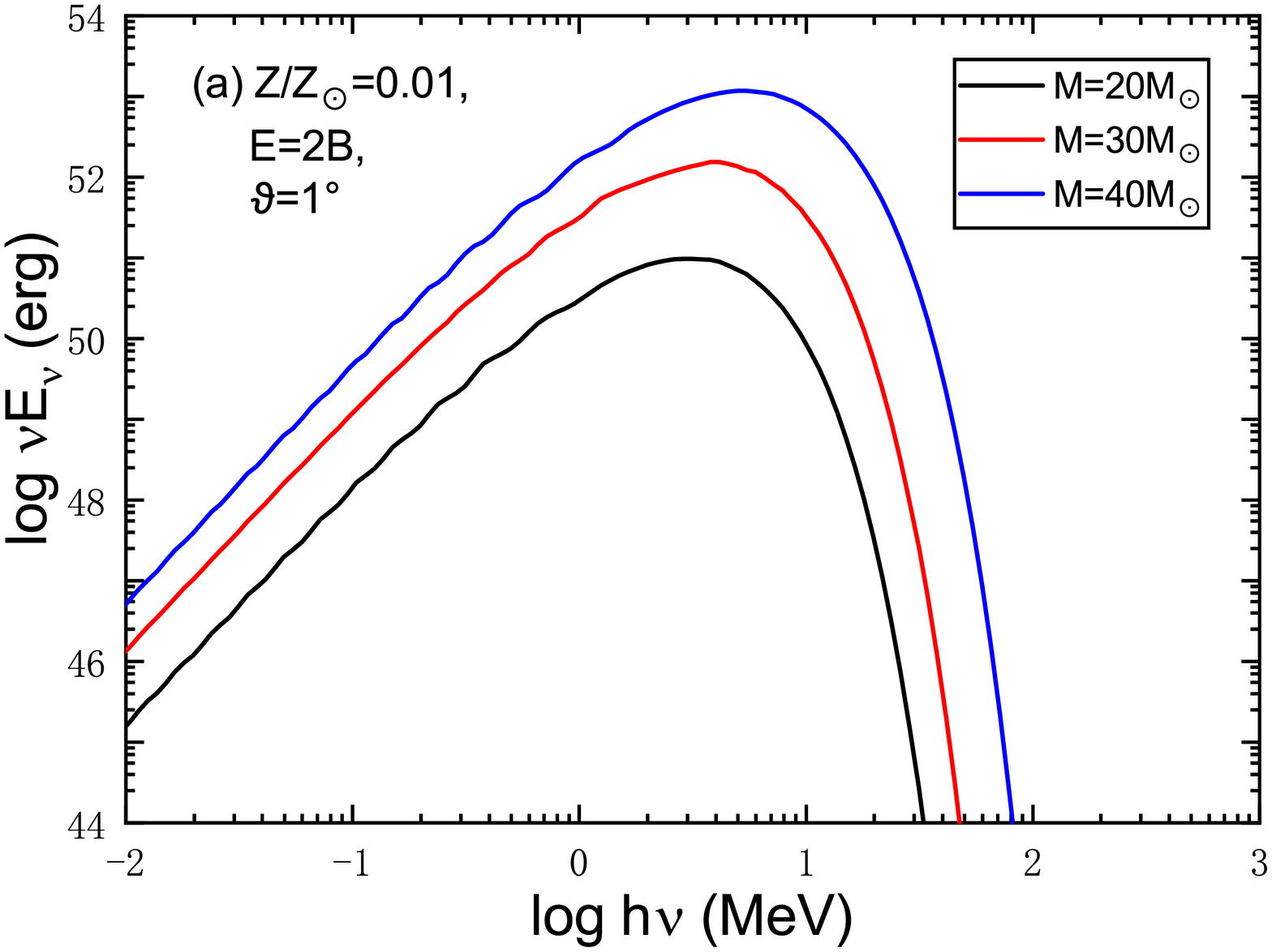}
\includegraphics[width=8cm,height=6cm]{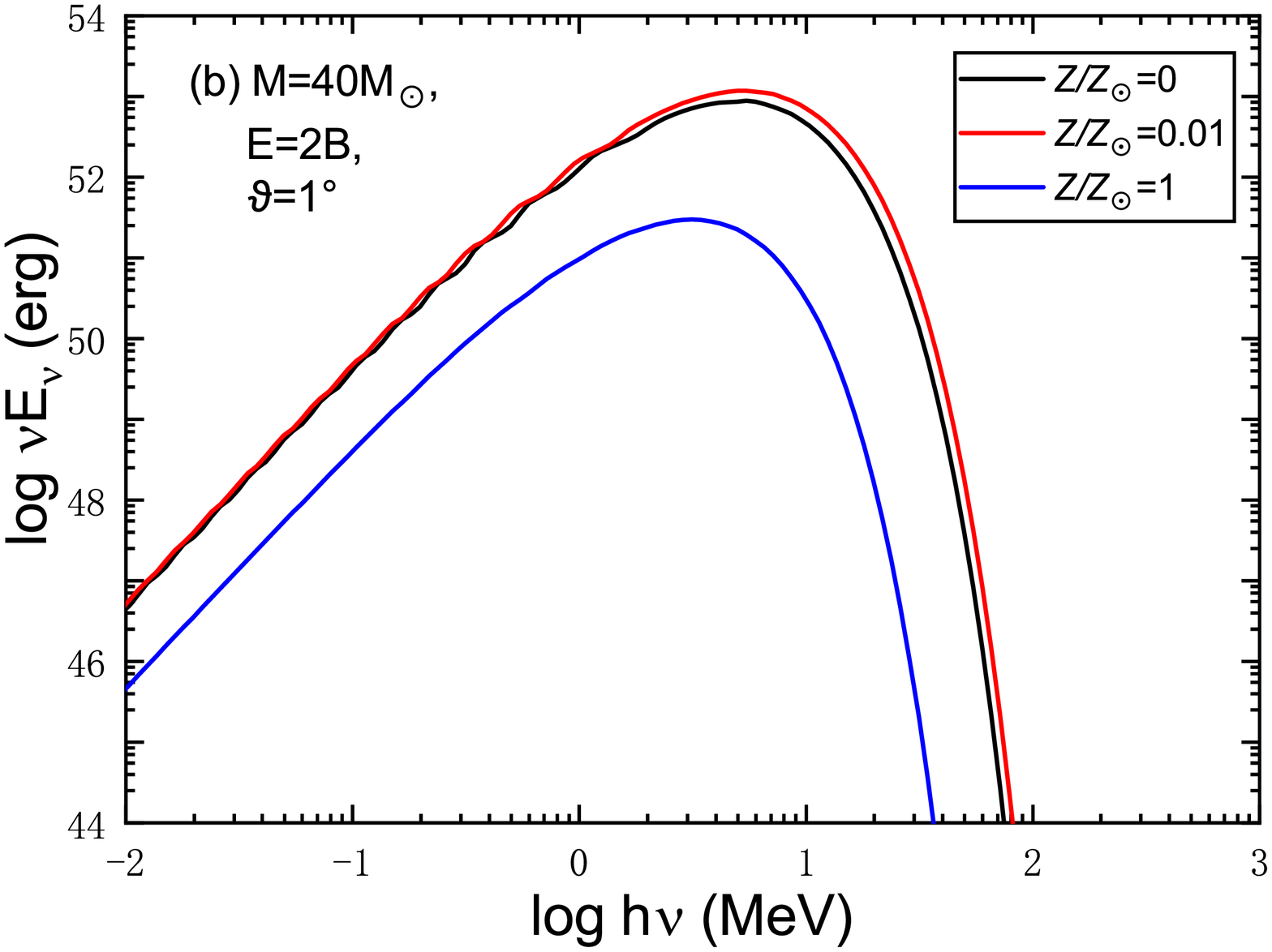}
\includegraphics[width=8cm,height=6cm]{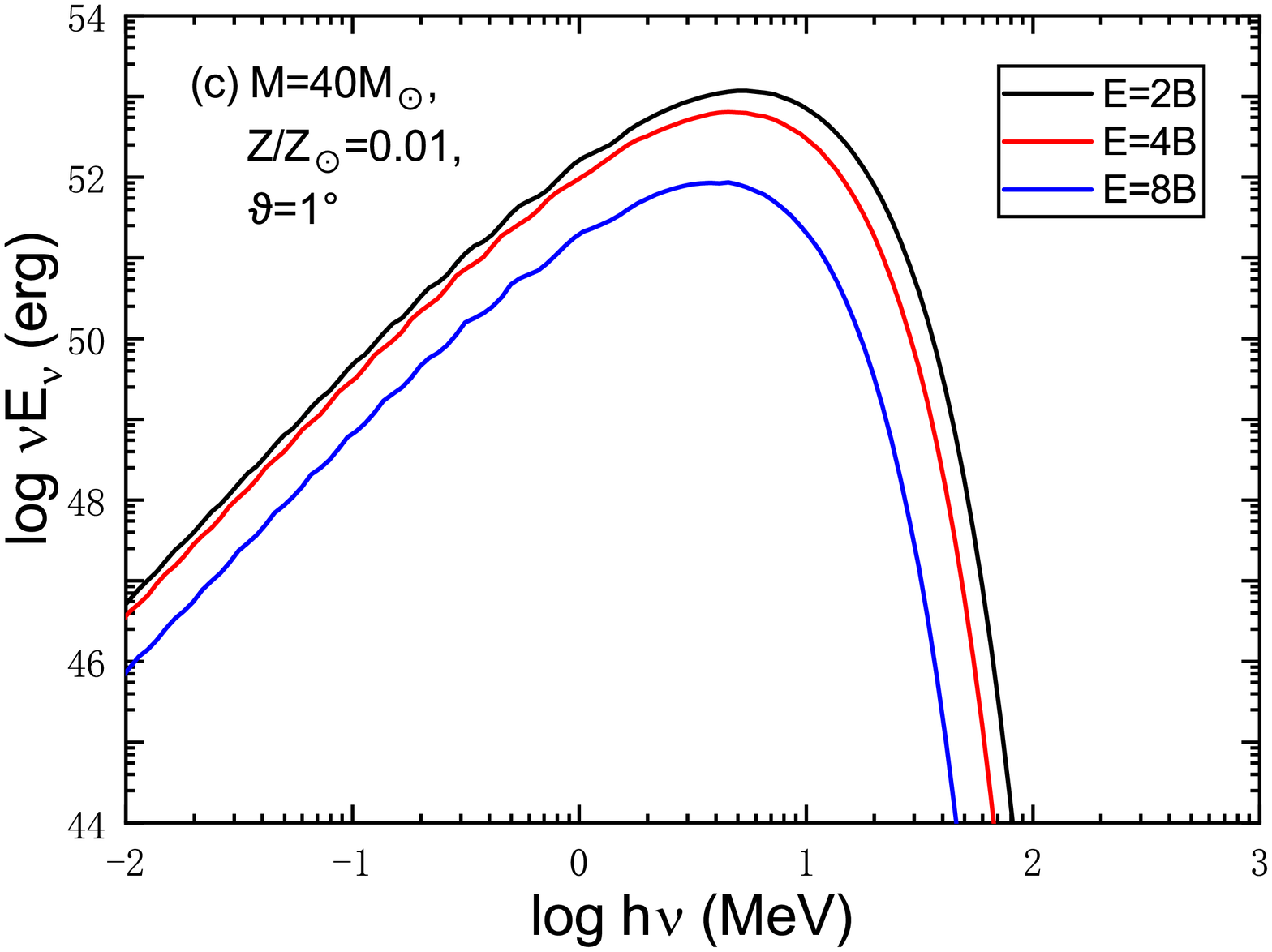}
\includegraphics[width=8cm,height=6cm]{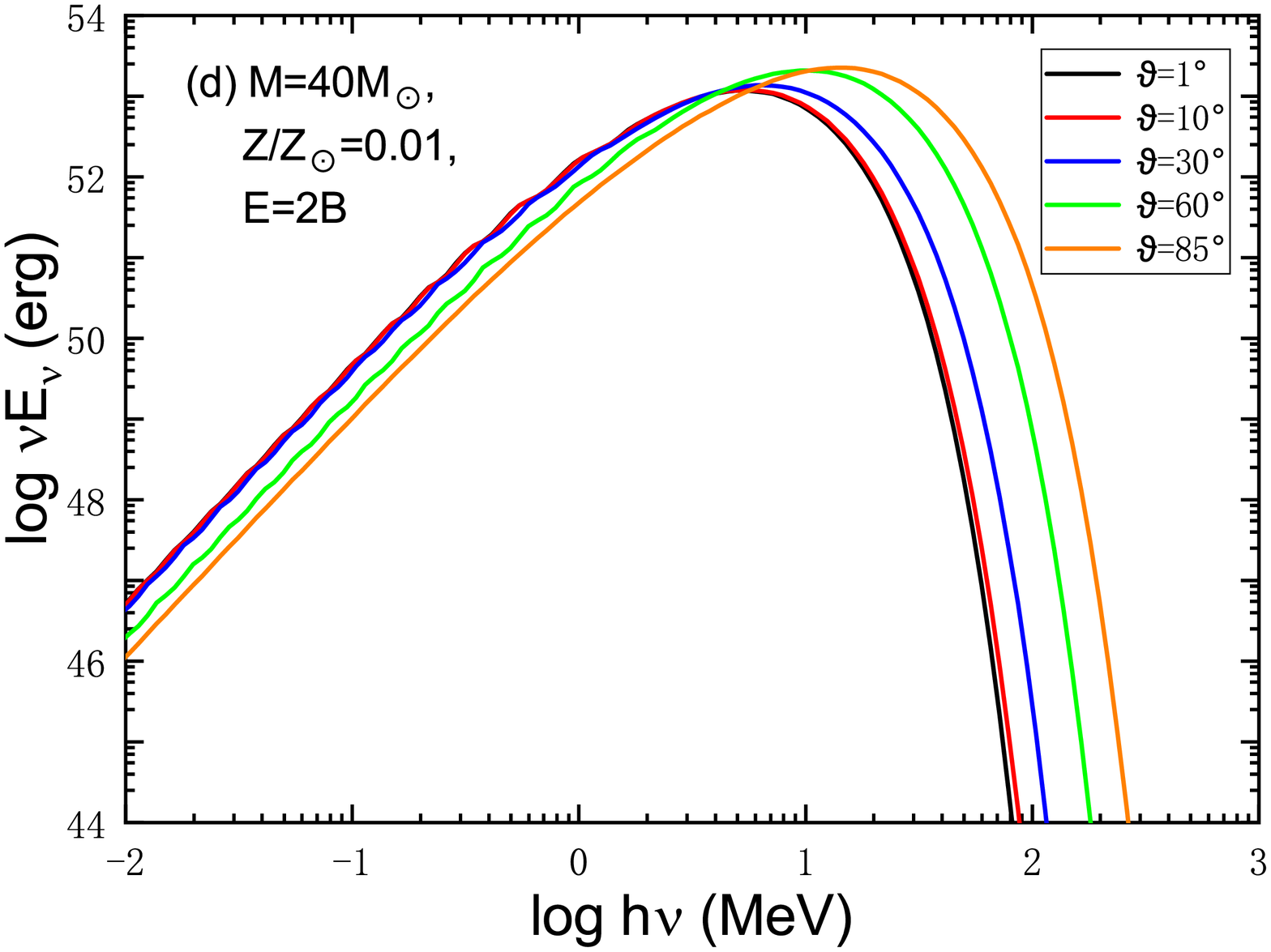}
\caption{Time-integrated electron antineutrino spectra of NDAFs as functions of the progenitor masses, metallicities, and initial explosion energies, and the viewing angles $\vartheta$.}
\label{fig2}
\end{figure*}

The mass and spin of a BH surrounded by a hyperaccretion disc will violently evolve within a short period. Based on the conversion of the energy and the angular momentum, the evolution equations of a spinning BH can be written as \citep[e.g.][]{Liu2012}
\beq
\frac{dM_{\rm{BH}}}{dt}=\dot{M}e_{\rm{ms}},
\eeq
\beq
\frac{dJ_{\rm{BH}}}{dt}=\dot{M}l_{\rm{ms}},
\eeq
where $M_{\rm{BH}}$, $J_{\rm{BH}}$, and $\dot{M}$ are the mass and angular momentum of the BH and the mass accretion rate, respectively. $e_{\rm{ms}}$ and $l_{\rm{ms}}$ are the specific energy and angular momentum at the marginally stable orbit, which are defined as \citep[see e.g.][]{Wu2013,Hou2014,Song2015}
\beq
e_{\rm{ms}}=\frac{1}{\sqrt{3x_{\rm{ms}}}}\left ( 4-\frac{3a_{*}}{\sqrt{x_{\rm{ms}}}} \right ),
\eeq
\beq
l_{\rm{ms}}=\frac{2\sqrt{3}GM_{\rm{BH}}}{c}\left ( 1-\frac{2a_{*}}{3\sqrt{x_{\rm{ms}}}} \right ),
\eeq
where $a_{*}\equiv cJ_{\rm{BH}}/GM_{\rm{BH}}^{2}$ is the dimensionless spin parameter of the BH. $x_{\rm{ms}}=3+Z_{2}-\sqrt{(3-Z_{1})(3+Z_{1}+2Z_{2})}$ is the dimensionless radius of the marginally stable orbit \citep{Bardeen1972,Kato2008}, where $Z_{1}=1+(1-a_{*}^{2})^{1/3}[(1+a_{*})^{1/3}+(1-a_{*})^{1/3}]$ and $Z_{2}=\sqrt{3_{*}^{2}+Z_{1}^{2}}$ for $0< a_{*}< 1$.
According to Equations (1)-(4), the evolution of the BH spin can be given by
\beq
\frac{da_{*}}{dt}=\frac{2\sqrt{3}\dot{M}}{M_{\rm{BH}}}\left ( 1- \frac{a_{*}}{\sqrt{x_{\rm{ms}}}}\right )^{2}.
\eeq

The effects of the initial explosion energies, the masses and metallicities of the massive progenitor stars on the time-evolution of the BH surrounded by the hyperaccretion disc are shown in Figure 1. The starting time is set at the time when the initial BH mass (core mass) is 2.3 $M_{\odot}$. For all cases, the initial BH spin parameter is set as $a_{*}=0.9$ in our calculations.

\section{Neutrinos from NDAFs}
\subsection{Model}

Based on the global solutions of NDAFs in \citet{Xue2013}, we derive the fitting formulae for the mean cooling rate due to electron antineutrino losses, $Q_{\bar{\nu}_{\rm{e}}}$ in units of $\rm erg~cm^{-2}~s^{-1}$, the temperature of the disc $T$ in units of $\rm K$, and the neutrino luminosity $L_{\rm{\nu }}$ in units of $\rm erg~s^{-1}$ as a function of the BH mass (in the range of $2.5-10~M_\odot$) and spin parameter, the mass accretion rate (less than $1~M_\odot~\rm s^{-1}$), and the radius, i.e.,
\begin{align}
\log Q_{\bar{\nu}_{\rm{e}}}=\ &41.40-0.23m_{\rm{BH}}+0.58a_{*}+1.85\log \dot{m}
\notag
\\&-3.96\log r,
\end{align}
\begin{align}
\log T=\ &11.23-0.4m_{\rm{BH}}+0.10a_{*}+0.23\log \dot{m}
\notag
\\&-0.86\log r,
\end{align}
\begin{align}
\log L_{\rm{\nu }}=\ 52.80-0.03 m_{\rm{BH}}+1.01a_{*}+1.08\log \dot{m},
\end{align}
where $m_{\rm{BH}}=M_{\rm{BH}}/M_{\odot}$ is the dimensionless BH mass, and $\dot{m}=\dot{M}/M_\odot~\rm s^{-1}$ and $r=R/R_g$ are the dimensionless mass accretion rate and radius, respectively. $R_g=2GM_{\rm{BH}}/c^2$ is the Schwarzschild radius.

The tracks of the neutrinos escaped from NDAFs should be effected by the central BHs. We use the well-known ray-tracing method \citep[e.g.][]{Fanton1997,Li2005,Liu2016} to calculate the neutrino propagation in a manner similar to the photon propagation near an accreting BH. For a given pixel of the image, the position of the emitter on the disc can be traced based on the null geodesic equation \citep{Carter1968}. By taking into account the corresponding gravitational potential and the velocity of the emission locations, the energy shift of a neutrino can be calculated. By integrating over all the pixels, the total observed flux distribution can be expressed as
\beq
F_{{E}_{\rm{obs}}}=\int_{\rm image}g^{3}I_{E_{\rm{em}}}d\Omega _{\rm{obs}},
\eeq
where $E_{\rm{obs}}$ is the observed neutrino energy, $E_{\rm{em}}$ is the neutrino emission energy from the local disc, $g \equiv E_{\rm obs} / E_{\rm em}$ is the energy shift factor, and $\Omega _{\rm{obs}}$ is the solid angle of the disc image to the observer. $I_{E_{\rm{em}}}$ is the local emissivity, which can be calculated according to the cooling rate $Q_{\bar{\nu}_{\rm{e}}}$ as
\beq
I_{E_{\rm{em}}}=Q_{\bar{\nu}_{\rm{e}}}\frac{F_{E_{\rm{em}}}}{\int F_{E_{\rm{em}}}dE_{\rm{em}}},
\eeq
where $F_{E_{\rm{em}}}=E_{\rm{em}}^{2}/[\exp (E_{\rm{em}}/kT-\eta )+1]$ is the unnormalized Fermi-Dirac spectrum \citep[e.g.][]{Rauch1994,Fanton1997,Li2005,Liu2016}.

\begin{figure}
\centering
\includegraphics[width=0.9\linewidth]{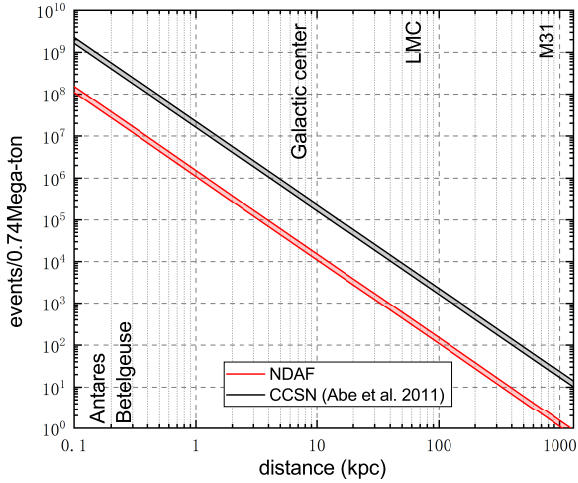}
\caption{Expected number of IBD events detected by Hyper-K for CCSNe and NDAFs as a function of the distance.}
\label{figure3}
\end{figure}

\subsection{Results}

Figure 2 shows the time-integrated spectra of the electron antineutrinos from NDAFs by considering the effects of the initial explosion energies, the viewing angles, and the masses and metallicities of the progenitor stars. The neutrino energies are generally in the range of 10 keV to 100 MeV and the peaks of the spectra occur at approximately $3-20$ MeV. For the same initial explosion energy, the higher progenitor mass and the lower metallicity are favourable for the neutrino emission of the disc. According to Equation (6), the neutrino-cooling rate decreases with increasing radius, which indicates that the high-energy neutrinos are mainly emitted from the inner region of the NDAF. These neutrinos are close to the BH and would be affected by the general relativistic effects. When the viewing angle $\vartheta$ increases, an increasing number of high-energy neutrinos deflected by the BH would be detected. On the other hand, the low-energy neutrinos are mostly emitted from the outer region of the disc, so that the general relativistic effects have little influence on them. Therefore, as shown in Figure 2(d), the viewing angle has more significant effects on the high-energy range of spectra. As the initial explosion energy decreases, the neutrino luminosity of NDAFs increases. The fallback mass supply rates are adopted, which are typically lower than the mass supply rates in the freefall approximation, and we consider the evolution of the mass and spin of the central BH, so the profiles of the neutrino spectra are different from the results of \citet{Wei2019}.

\begin{figure}
\centering
\includegraphics[width=1.0\linewidth]{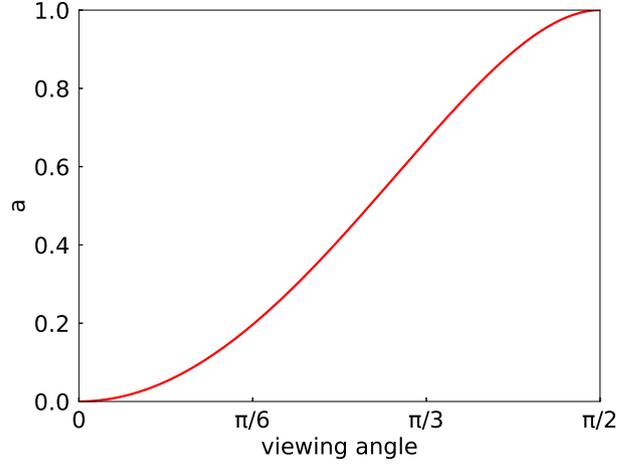}
\caption{The dependence of the GW amplitude on the viewing angle $\vartheta$.}
\label{figure4}
\end{figure}

Electron antineutrinos can be detected via inverse beta decay (IBD) reaction $\bar{\nu} _{\rm{e}}+p\rightarrow n+e^{+}$ by the upcoming Hyper-Kamiokande (Hyper-K) detector. Neutrinos of different flavours are supposed to be detected by Hyper-K, but the IBD is the best detection channel \citep{Abe2011}. The expected number of the NDAF neutrino events detected by Hyper-K can be estimated in reference to CCSN neutrinos. The CCSN neutrino emission has been widely investigated \citep[e.g.][]{Hudepohl2010,Scholberg2012,OConnor2013,Seadrow2018,Takiwaki2018,Glas2019,Li2019,Li2021,Mueller2019,Vartanyan2019,Morinaga2020,Walk2020,Warren2020,Nagakura2021a,Nagakura2021b,Suwa2019,Suwa2021}. \citet{Nagakura2021a,Nagakura2021b} studied the CCSN neutrino signals and gave the event rates, the observed energy spectra, and the cumulative number of the events of some terrestrial neutrino detectors. They also found that there is a correlation between the total neutrino energy and the cumulative number of the neutrino events for the detectors. For the typical energy released by a CCSN is about $\sim 3\times 10^{53}$ erg, almost 99$\%$ of the released energy is carried out by neutrinos. Then approximately $165,000-230,000$ inverse beta events are expected to be detected by Hyper-K for a typical CCSN at a distance of 10 kpc \citep{Abe2011}. Figure 3 shows the expected number of the IBD events for CCSNe and NDAFs by Hyper-K. Obviously, if we adopt the typical total neutrino energy of NDAFs as $\sim 2 \times 10^{52}$ erg, the typical events of CCSNe are about one order of magnitude higher than those of NDAFs. In the case of the Large Magellanic Cloud (LMC) where SN 1987a was located, approximately several hundred events are expected for NDAFs. At the distance of M31 (Andromeda Galaxy), only a few NDAF neutrinos can be detected. The detailed detection rate for NDAFs in the Local Group was estimated in \citet{Liu2016}. In the Local Group, if one takes the event rate of the SN Ib/c as an optimistic event rate for NDAFs, the expected detection rate for NDAFs is $1-3$ per century for the Hyper-K detector. Other neutrino detectors such as the Jiangmen Underground Neutrino Observatory \citep{An2016}, Super-Kamiokande \citep{Abe2014}, and LENA \citep{Wurm2012} can also detect NDAF neutrinos, but the detection distance is more limited.

\section{GWs from NDAFs}
\subsection{Model}

\begin{figure}
\centering
\includegraphics[width=8cm,height=6cm]{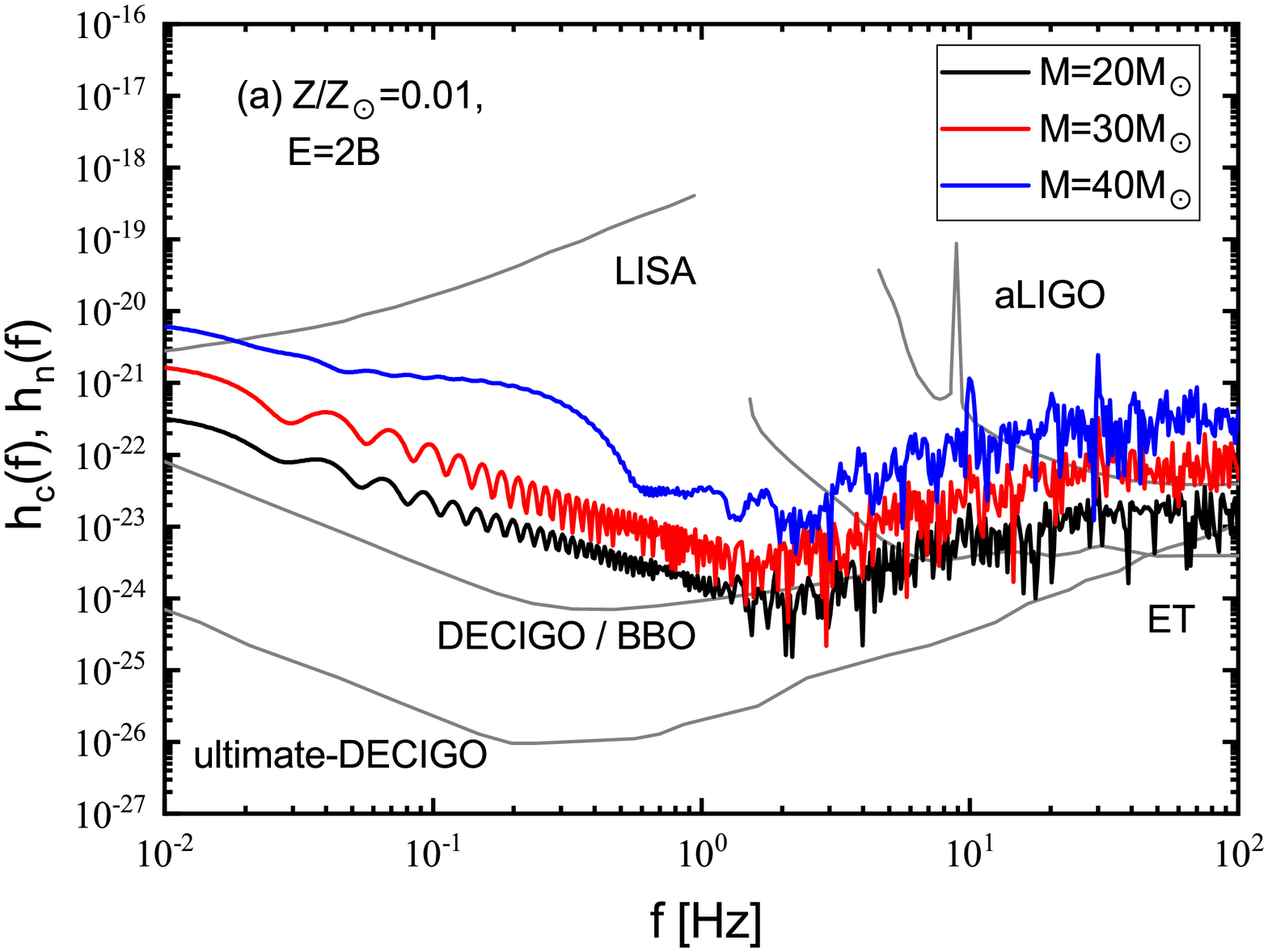}
\includegraphics[width=8cm,height=6cm]{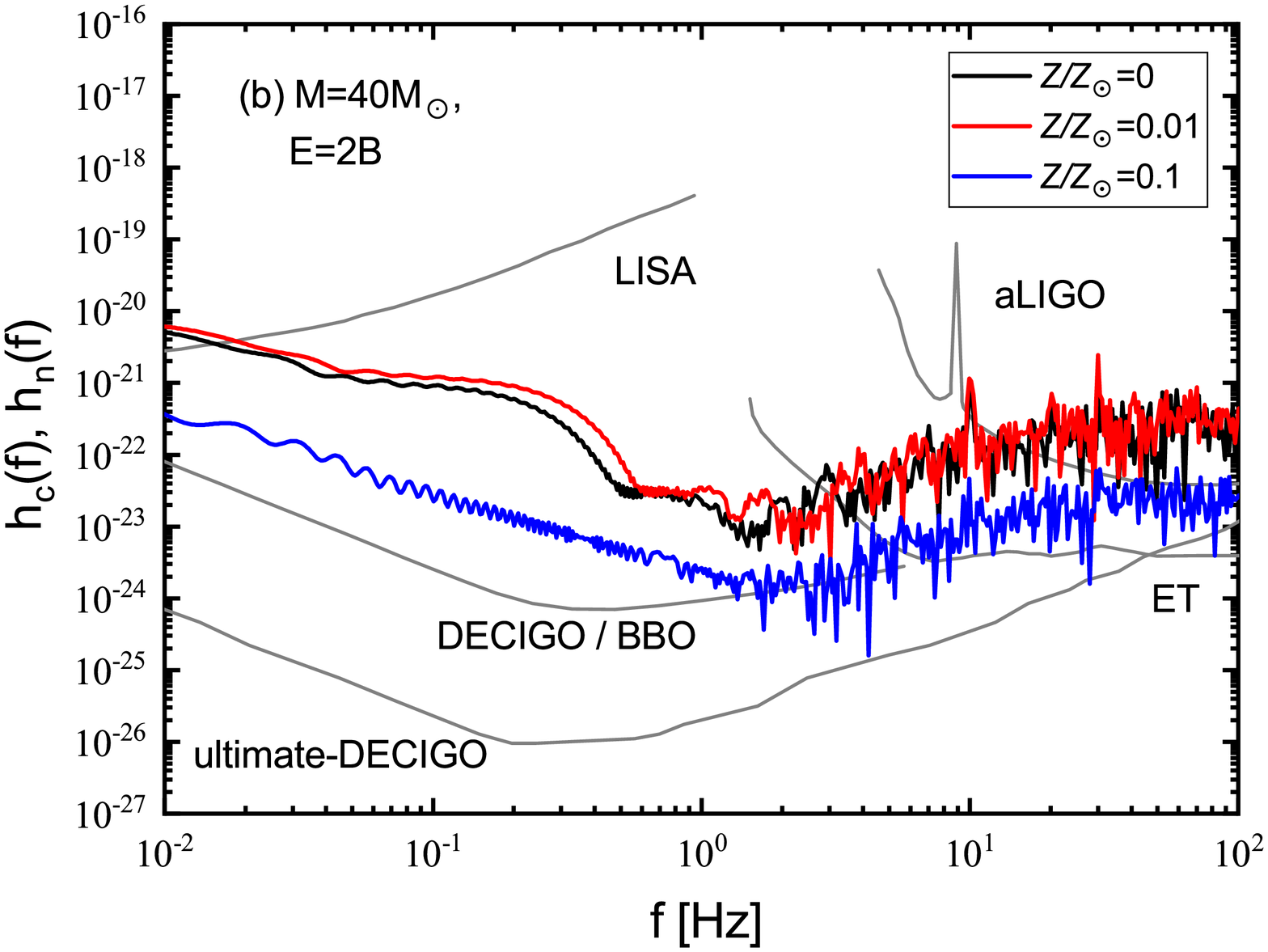}
\includegraphics[width=8cm,height=6cm]{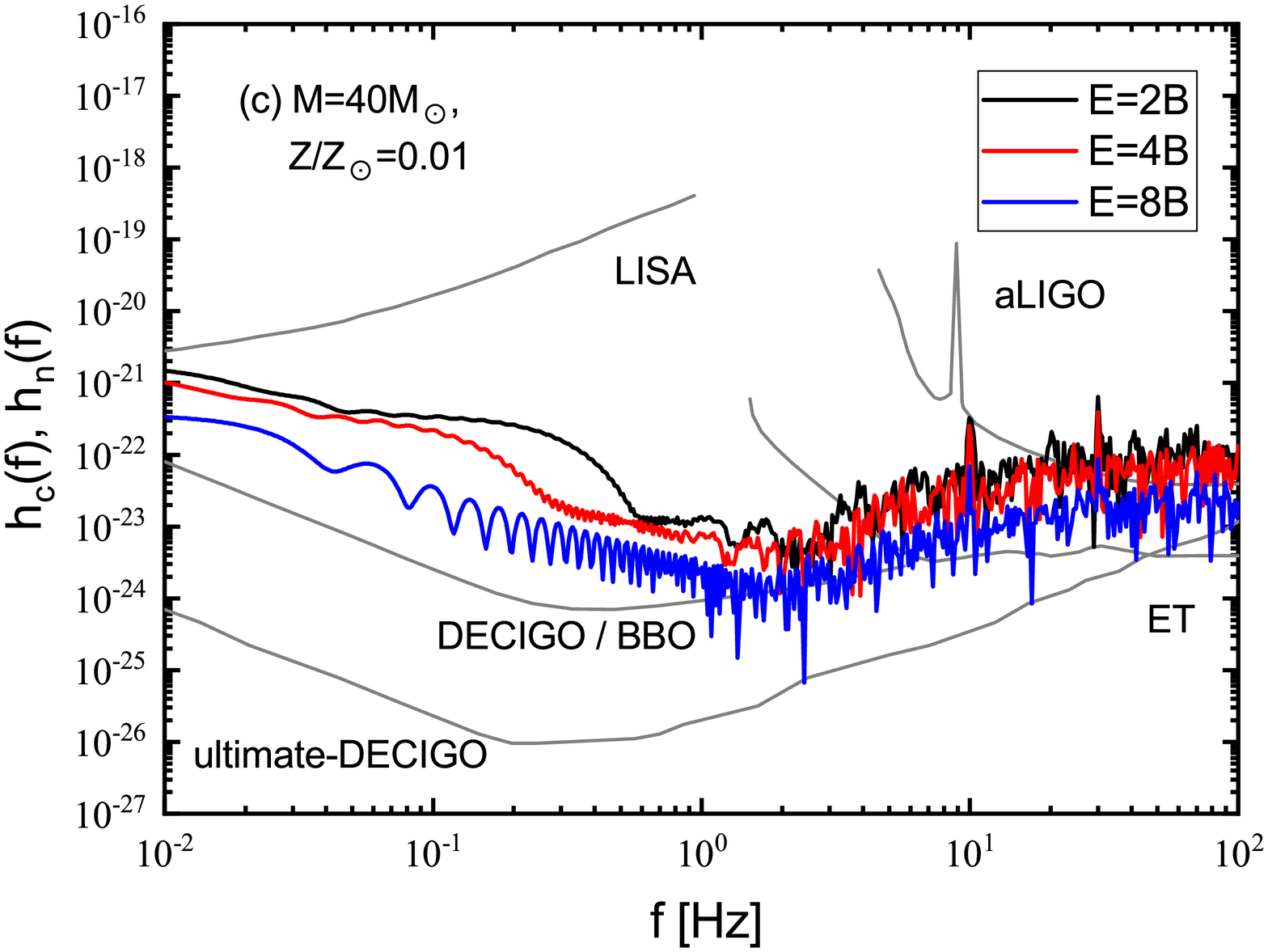}
\caption{Characteristic amplitudes of GWs from NDAFs as functions of progenitor masses, metallicities and initial explosion energies. }
\label{fig5}
\end{figure}

\citet{Epstein1978} first analyzed the GWs from a small source due to the anisotropic axisymmetric emission of neutrinos. The neutrino-induced GWs from CCSNe were investigated in some simulations \citep[see e.g.][]{Burrows1996,Mueller1997,Kotake2006,Kotake2007,Kotake2009,Mueller2012,Mueller2013,Vartanyan2020}. \citet{Vartanyan2020} investigated GWs from CCSNe sourced by neutrino emission asymmetries for a wide range of progenitor masses and concluded that they could be detected at 10 kpc by aLIGO, ET, and DECIGO. Besides, neutrino-induced GWs from NDAFs were studied in some previous literatures \citep{Suwa2009,Liu2017b}. The formulas of the GW amplitude for the axisymmetric emission of neutrinos from NDAFs have been introduced in \citet{Suwa2009} and \citet{Wei2020}. We simplify the NDAF as a geometrically infinitely thin disc and assume that the emission of neutrinos is isotropic at the disc surface. The GW amplitude is then given by
\beq
h_{+}(t,\vartheta)=&&\frac{1+2 \cos\vartheta }{3}\tan^{2}(\frac{\vartheta }{2})\frac{2G}{R_{\rm obs} c^{4}} \nonumber \\&& \times \int_{-\infty }^{t-R_{\rm obs}/c}L_{\rm{\nu}}(t')dt',
\eeq
where $R_{\rm obs}$ is the distance from the observer to the source and $L_{\rm{\nu}}(t')dt'$ is the neutrino luminosity of the source. As shown in Equation (11), the GW amplitude depends on the viewing angle. We can define an angle factor $a=(1+2 \cos\vartheta)\tan^{2}(\vartheta /2)$. The effect of viewing angle on the GW amplitude is displayed in Figure 4. When the observer is located in the equatorial plane, i.e., $\vartheta=\pi/2$, the GW amplitude is the largest.

The local energy flux of GWs can be expressed as \citep[e.g.][]{Suwa2009}
\beq
\frac{dE_{\rm{GW}}}{R_{\rm obs}^{2}d\Omega dt}=\frac{c^{3}}{16\pi G}\left | \frac{d}{dt}h_{+}(t,\vartheta) \right |^{2},
\eeq
where $\Omega$ is the solid angle in the observer coordinate frame.

Integrating over a sphere surrounding the source, the total energy can be obtained as
\beq
E_{\rm{GW}}=\frac{\beta G}{9c^{5}}\int_{-\infty }^{\infty }dtL_{\rm{\nu}}(t)^{2},
\eeq
where $\beta \sim 0.47039$. In order to acquire a GW spectrum, we write $L_{\rm{\nu}}(t)$ in terms of the inverse Fourier transform as
\beq
L_{\rm{\nu} }(t)=\int_{-\infty }^{+\infty} \tilde{L}_{\rm{\nu} }(f)e^{-2\pi ift}df,
\eeq
then, the GW energy spectrum can be expressed as
\beq
\frac{dE_{\rm{GW}}(f)}{df}=\frac{2\beta G}{9c^{5}}\left |  \tilde{L}_{\rm{\nu} }(f)\right |^{2}.
\eeq

The characteristic GW strains are defined by
\beq
h_{\rm{c}}(f)=\frac{1}{R_{\rm obs}}\sqrt{\frac{2}{\pi ^{2}}\frac{G}{c^{2}}\frac{dE_{\rm{GW}}(f)}{df}}
\eeq
for a given frequency $f$ \citep[e.g.][]{Flanagan1998}.

Moreover, we can calculate signal-to-noise ratios (SNRs) obtained from matched filtering in the GW detectors. The optimal SNR is expressed as
\beq
{\rm{SNR^{2}}}=\int_{0}^{\infty}d(\ln{f})\frac{h_{\rm{c}}(\emph{f})^{2}}{h_{\rm{n}}(\emph{f})^{2}},
\eeq
where $h_{\rm{n}}f=\sqrt{5fS_{\rm{h}}(f)}$ is the noise amplitude and $S_{\rm{h}}(f)$ is the spectral density of the strain noise in the detector at frequency $f$.

\subsection{Results}

Figure 5 shows the strains of the GWs from NDAFs at a distance of 10 kpc. The typical frequency of GWs from NDAFs is $1-100$ Hz. The gray lines represent the sensitivity curves (the noise amplitudes $h_{\rm{n}}$) of aLIGO, ET, LISA, DECIGO/BBO, and ultimate-DECIGO. At a distance of 10 kpc, the neutrino-induced GWs can be detected by aLIGO and ET in the frequency range of $\sim$ $10-100$ Hz and be detected by DECIGO/BBO and ultimate-DECIGO in the frequency range of $\sim$ $1-10$ Hz. As shown in Equation (11), the GW amplitude is determined by the neutrino luminosity $L_{\nu}$, and hence, the GW emissions are related to the mass accretion rate. Therefore, the effects of the properties of progenitor stars and the initial explosion energy on the GW strains are similar to those of the factors on the neutrino spectra. Higher progenitor mass, lower metallicity and lower initial explosion energy are favourable for the GW emission of NDAFs. In contrast to the results of \citet{Wei2020}, the profiles of the GW spectra exhibit differences. This is because the mass accretion rates are obtained in the different ways and the evolution of the central BH is considered.

\section{Summary}

In this work, we adopt pre-SN models and simulate a series of CCSN explosions. Based on these calculations, we obtain the fallback rates from different progenitors with some specific initial explosion energies. We set the fallback mass supply rate as the mass accretion rate of the disc. By considering the evolution of the BH during the hyperaccretion process, we calculate the neutrino emission and GW radiation from NDAFs. These MeV neutrino signals and neutrino-induced GW signals are hopefully detected by the future detectors in the Local Group. Even if the centre does not form an accretion disc, the CCSN fallback would emit large amounts of MeV neutrinos \citep[e.g.][]{Fryer2009,Wei2019}. Therefore, the joint detections of neutrinos and GWs, including the observations of LGRBs associated with CCSN, are meaningful and would help verify the existence of the central BH hyperaccretion disc.

The effects of the initial explosion energies, viewing angles, and masses and metallicities of progenitor stars on time-integrated spectra of neutrinos and GW spectra are investigated. Higher progenitor mass, lower metallicity, and lower initial explosion energy are favourable for the neutrino emission and neutrino-induced GWs radiation of NDAFs. One can notice that the initial explosion energy in the range of $2-8$ B is relatively small. In \citet{Liu2021}, we displayed the residual explosion energy of all cases in Table 1. For the progenitor stars ($\sim 20-40~M_\odot$) with zero-metallicity, the residual energy is in the range of $0.01-1.01$ B for the cases with the initial explosion energy $\sim 2$ B, and about 6 B for 8 B. It is expected that only a small fraction of the residual energy can be converted into the radiation energy. According to the current CCSN observations, faint or failed CCSNe with low initial explosion energy might be universal, which is beneficial to the detection of the neutrino emission and GW radiation of NDAFs.

\section*{Acknowledgements}

This work was supported by the National Natural Science Foundation of China under grant 11822304, the science research grants from the China Manned Space Project with No. CMS-CSST-2021-B11, and the Natural Science Foundation of Fujian Province of China under grant 2018J01007.

\section*{Data availability}

The data underlying this article will be shared on reasonable request to the corresponding author.

\appendix
\section{Results of CCSN simulations}

In this appendix, we briefly present the results of one-dimensional CCSN simulations. The effects of the initial explosion energies, masses and metallicities of the progenitors on the evolution of explosions and fallbacks are exhibited as follows. The profiles of the densities and velocities at 50, 100, 500, and 1,000 s for the different explosions are shown in Figures A1a-A1c and Figures A2a-A2c, respectively. Figures A3a-A3c show the time evolutions of the mass supply rates for all simulations and all slopes of the fallback mass rates follow $\sim -5/3$ \citep{Liu2021}. In all figures, the signs a, b, and c correspond to the progenitor metallicities of $Z/Z_\odot$ = 0, 0.01, and 1, respectively.

\begin{figure*}
 \ContinuedFloat*
 \includegraphics[width=10cm,height=10cm]{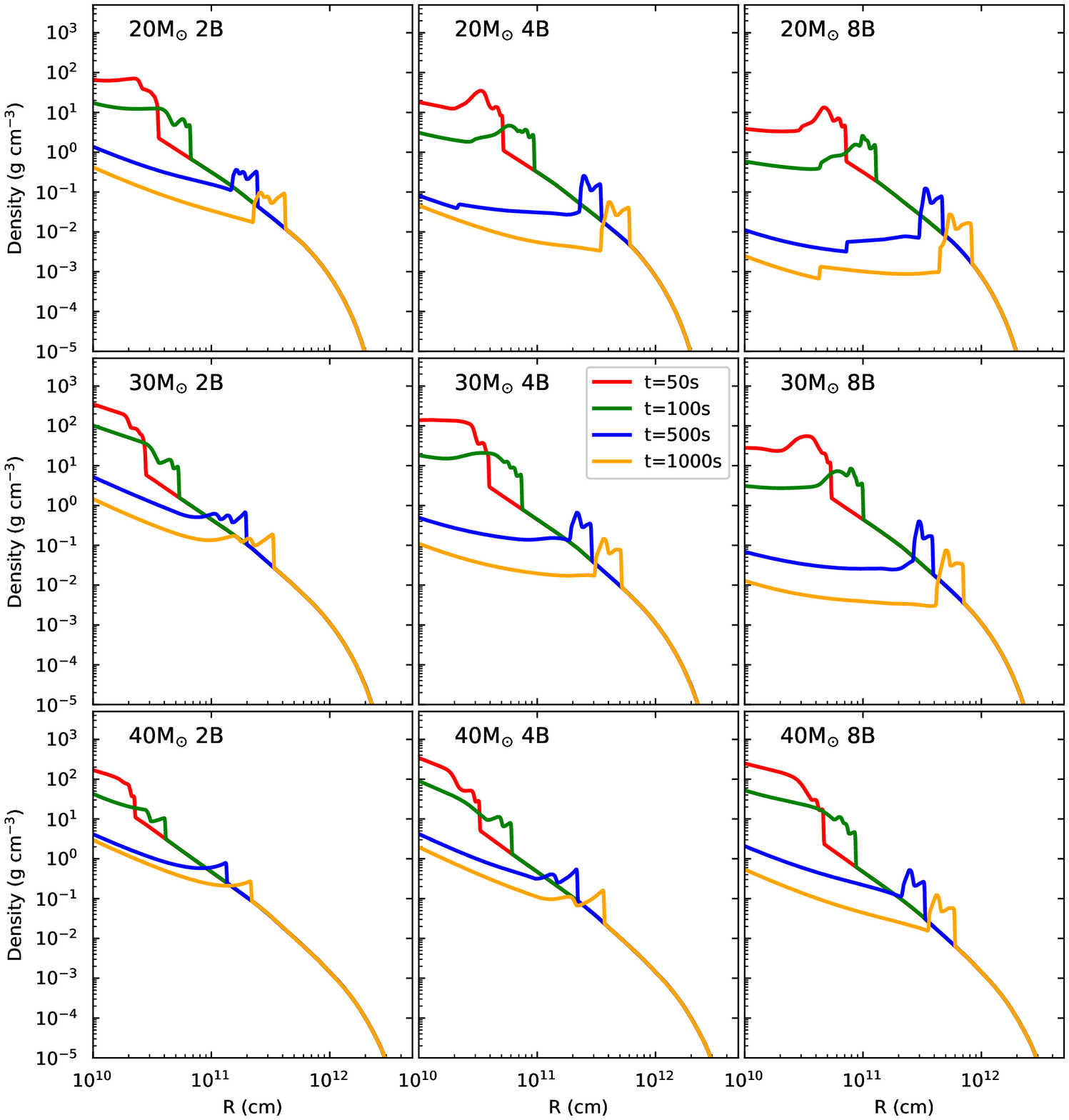}
 \caption{Profiles of densities at 50, 100, 500, and 1,000 s with the progenitor metallicity $Z/Z_\odot$ = 0.}
\end{figure*}

\begin{figure*}
 \ContinuedFloat
 \includegraphics[width=10cm,height=10cm]{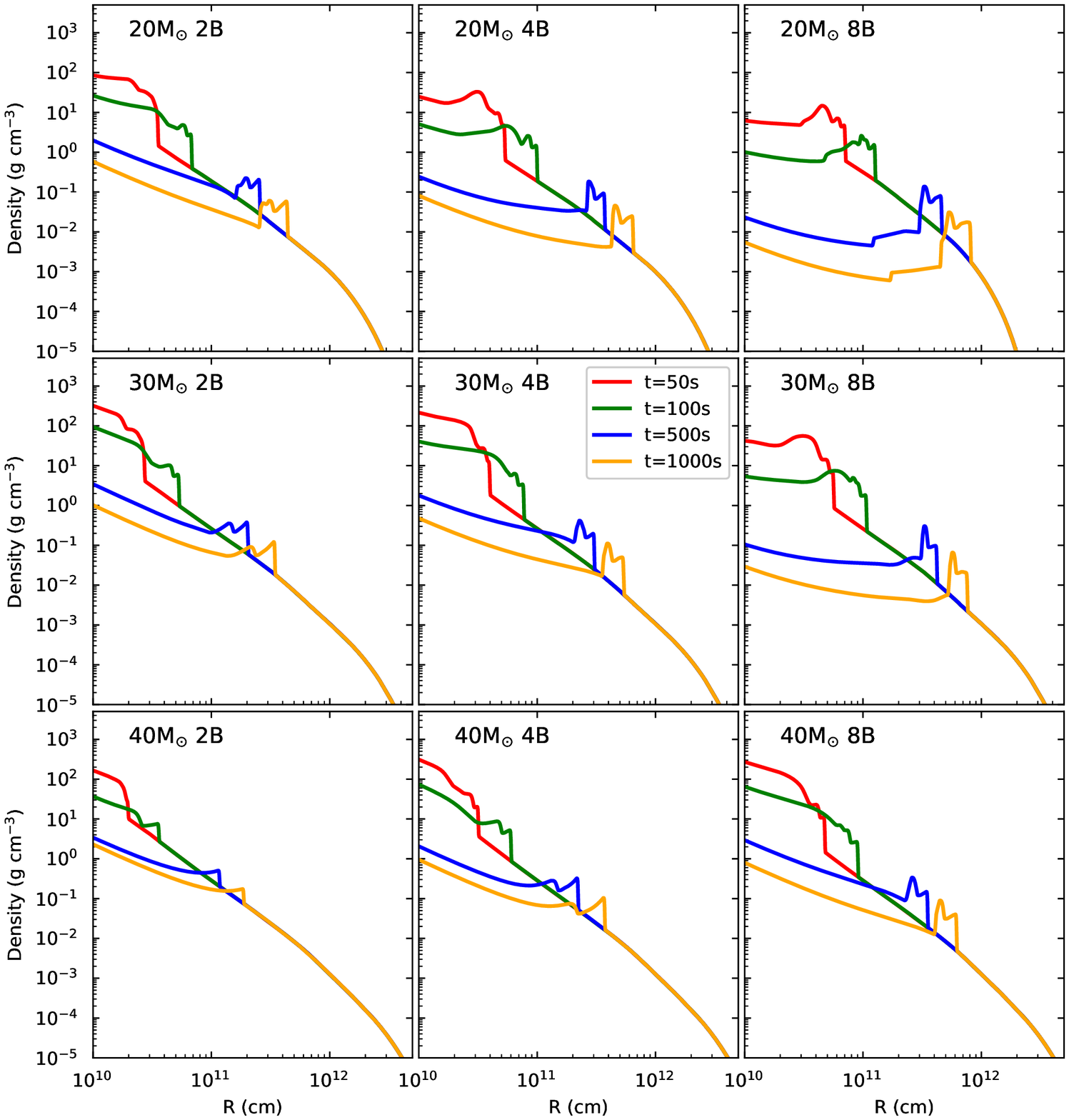}
 \caption{Profiles of densities at 50, 100, 500, and 1,000 s with the progenitor metallicity $Z/Z_\odot$ = 0.01.}
\end{figure*}

\begin{figure*}
 \ContinuedFloat
 \includegraphics[width=10cm,height=10cm]{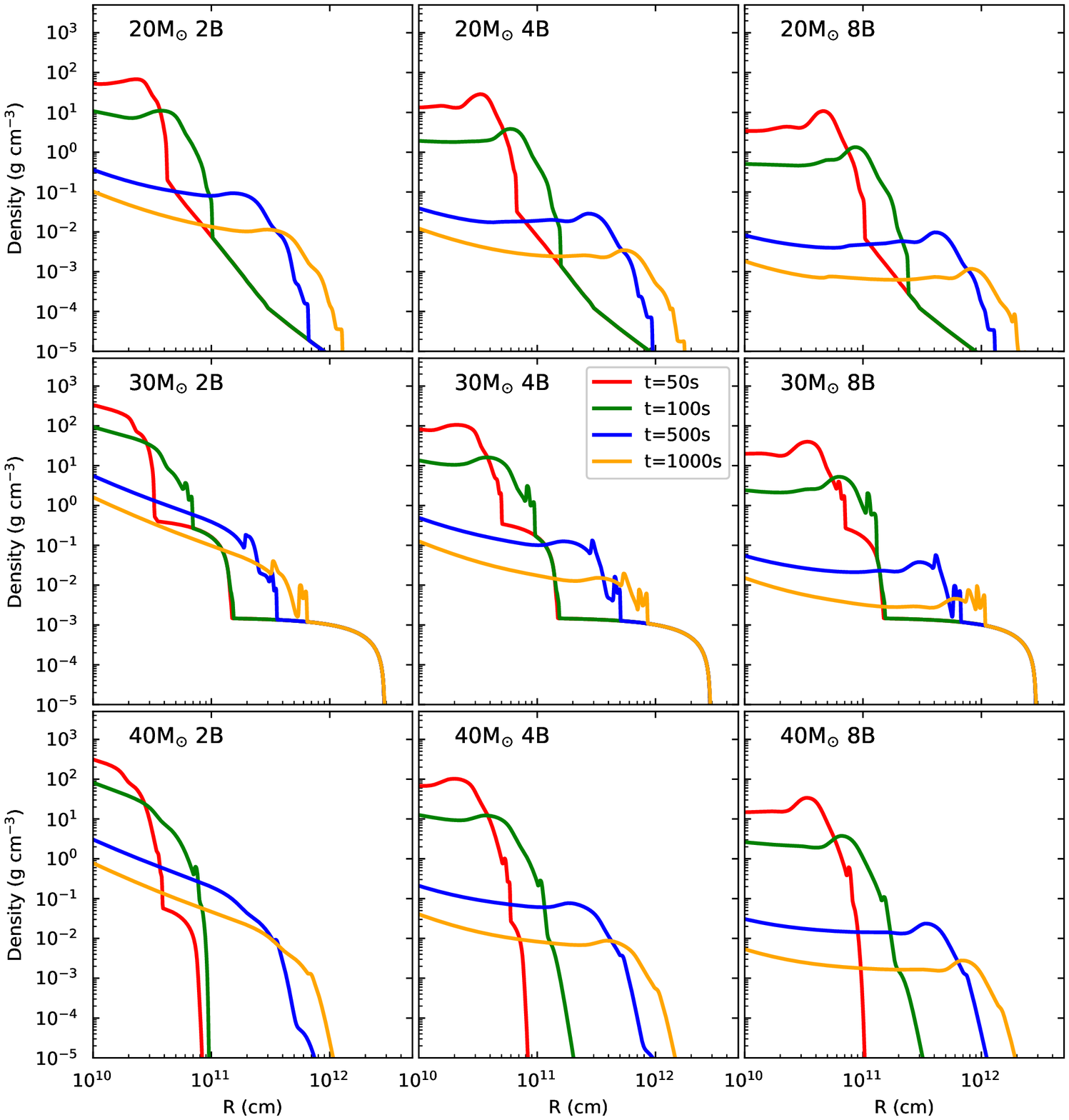}
 \caption{Profiles of densities at 50, 100, 500, and 1,000 s with the progenitor metallicity $Z/Z_\odot$ = 1.}
\end{figure*}

\begin{figure*}
 \ContinuedFloat*
 \includegraphics[width=10cm,height=10cm]{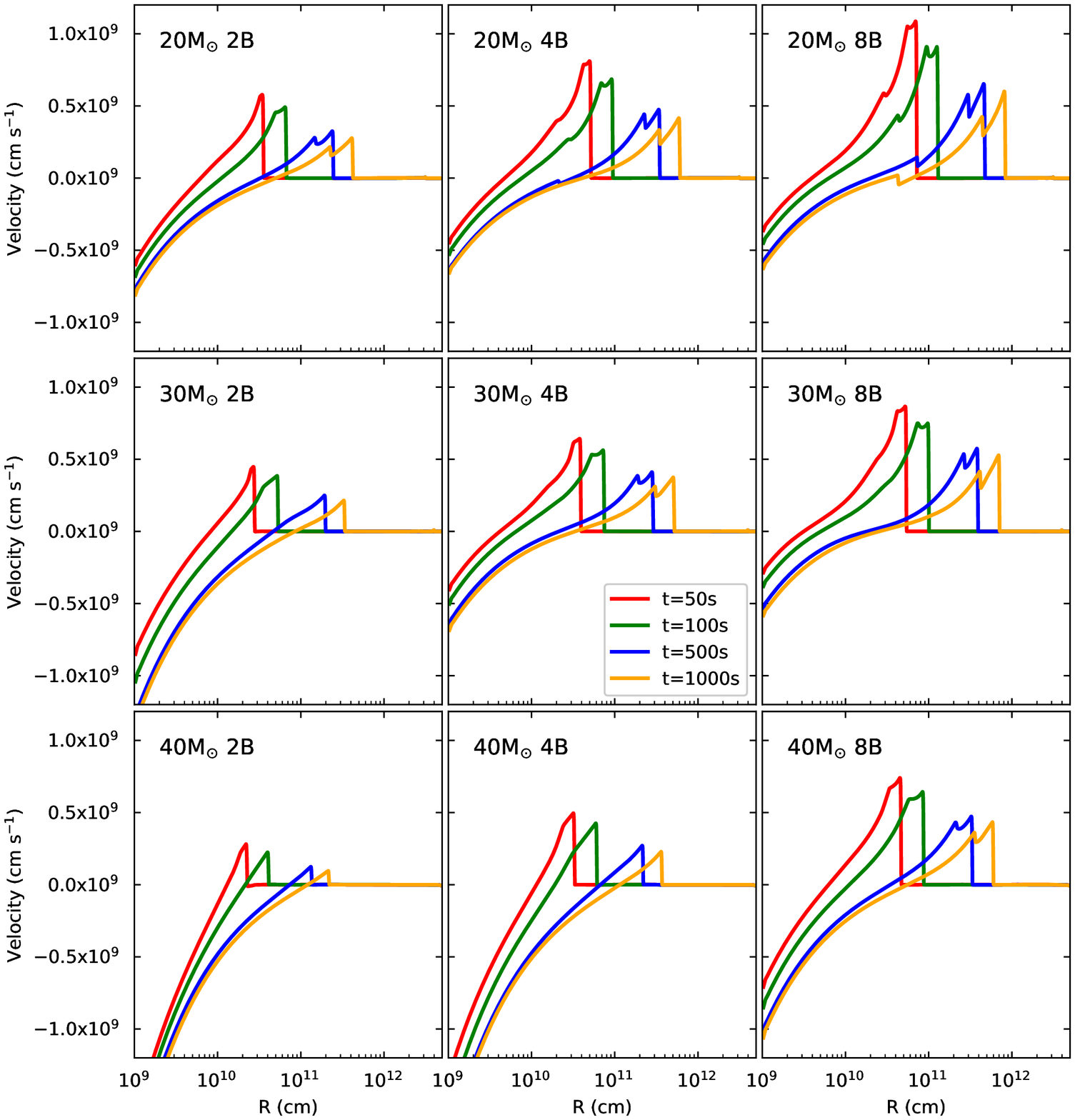}
 \caption{Profiles of velocities at 50, 100, 500, and 1,000 s with the progenitor metallicity $Z/Z_\odot$ = 0.}
\end{figure*}

\begin{figure*}
 \ContinuedFloat
 \includegraphics[width=10cm,height=10cm]{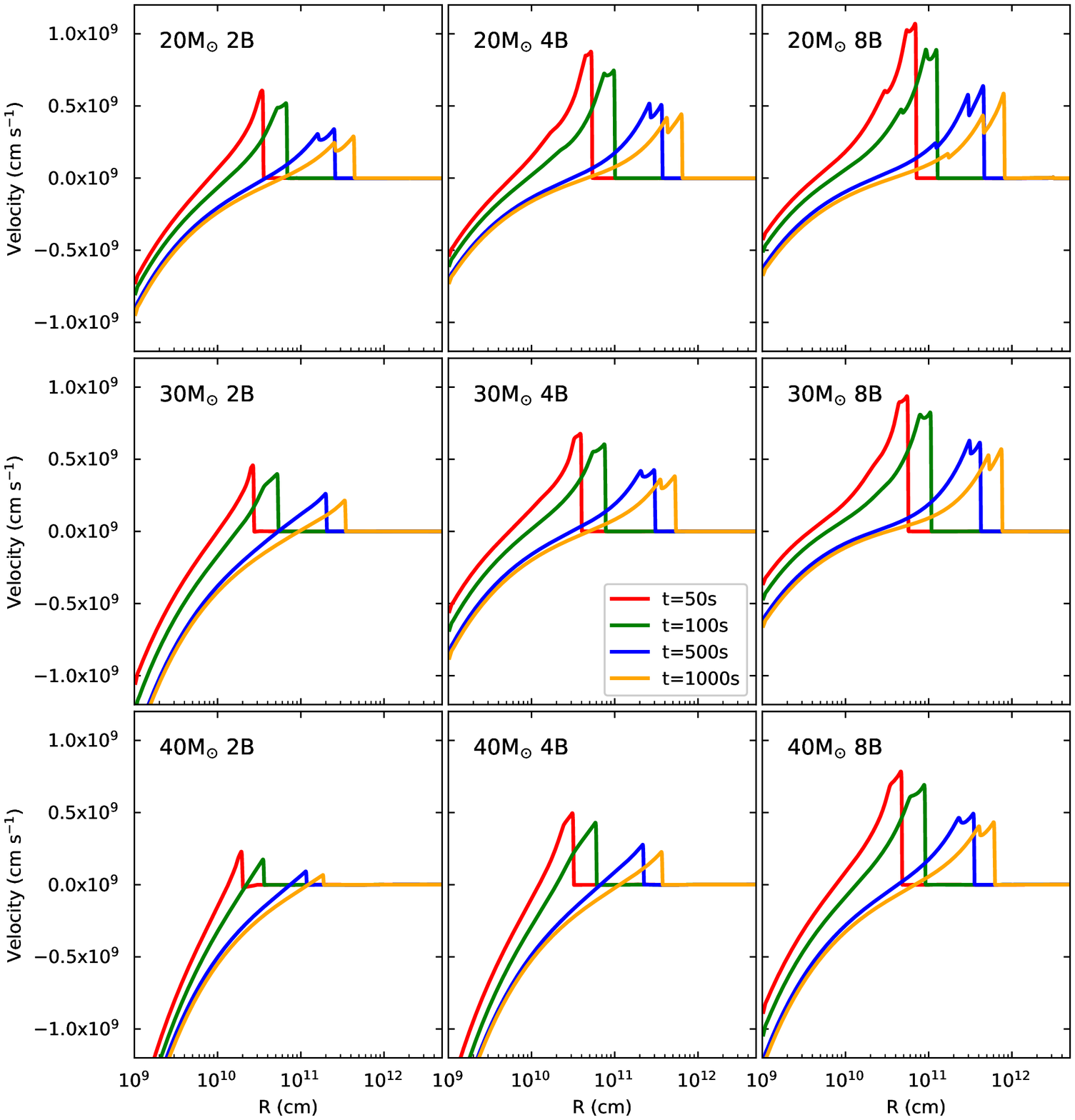}
 \caption{Profiles of velocities at 50, 100, 500, and 1,000 s with the progenitor metallicity $Z/Z_\odot$ = 0.01.}
\end{figure*}

\begin{figure*}
 \ContinuedFloat
 \includegraphics[width=10cm,height=10cm]{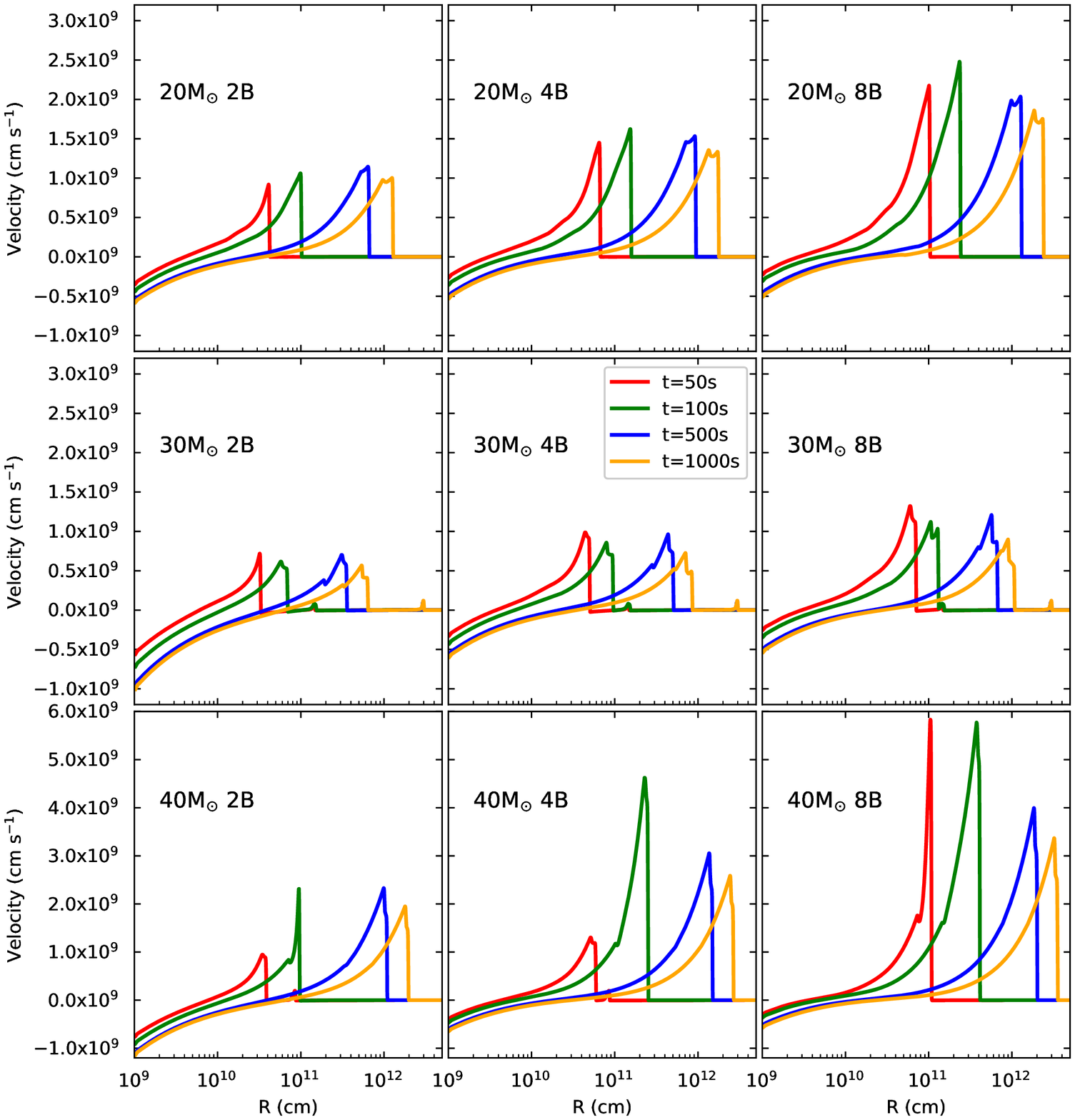}
 \caption{Profiles of velocities at 50, 100, 500, and 1,000 s with the progenitor metallicity $Z/Z_\odot$ = 1.}
\end{figure*}

\begin{figure*}
 \ContinuedFloat*
 \includegraphics[width=10cm,height=10cm]{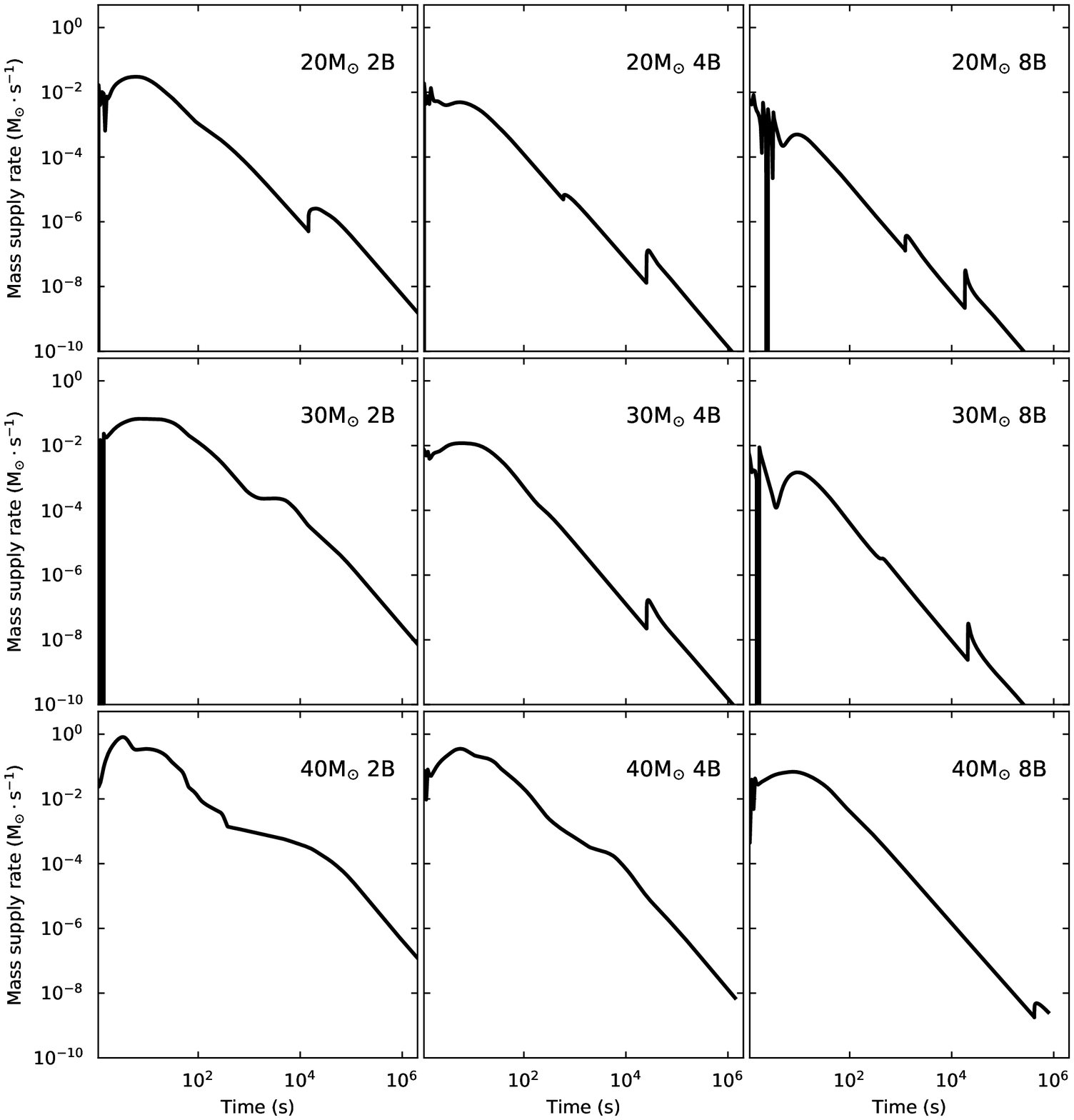}
 \caption{Time evolutions of mass supply rates with the progenitor metallicity $Z/Z_\odot$ = 0.}
\end{figure*}

\begin{figure*}
 \ContinuedFloat
 \includegraphics[width=10cm,height=10cm]{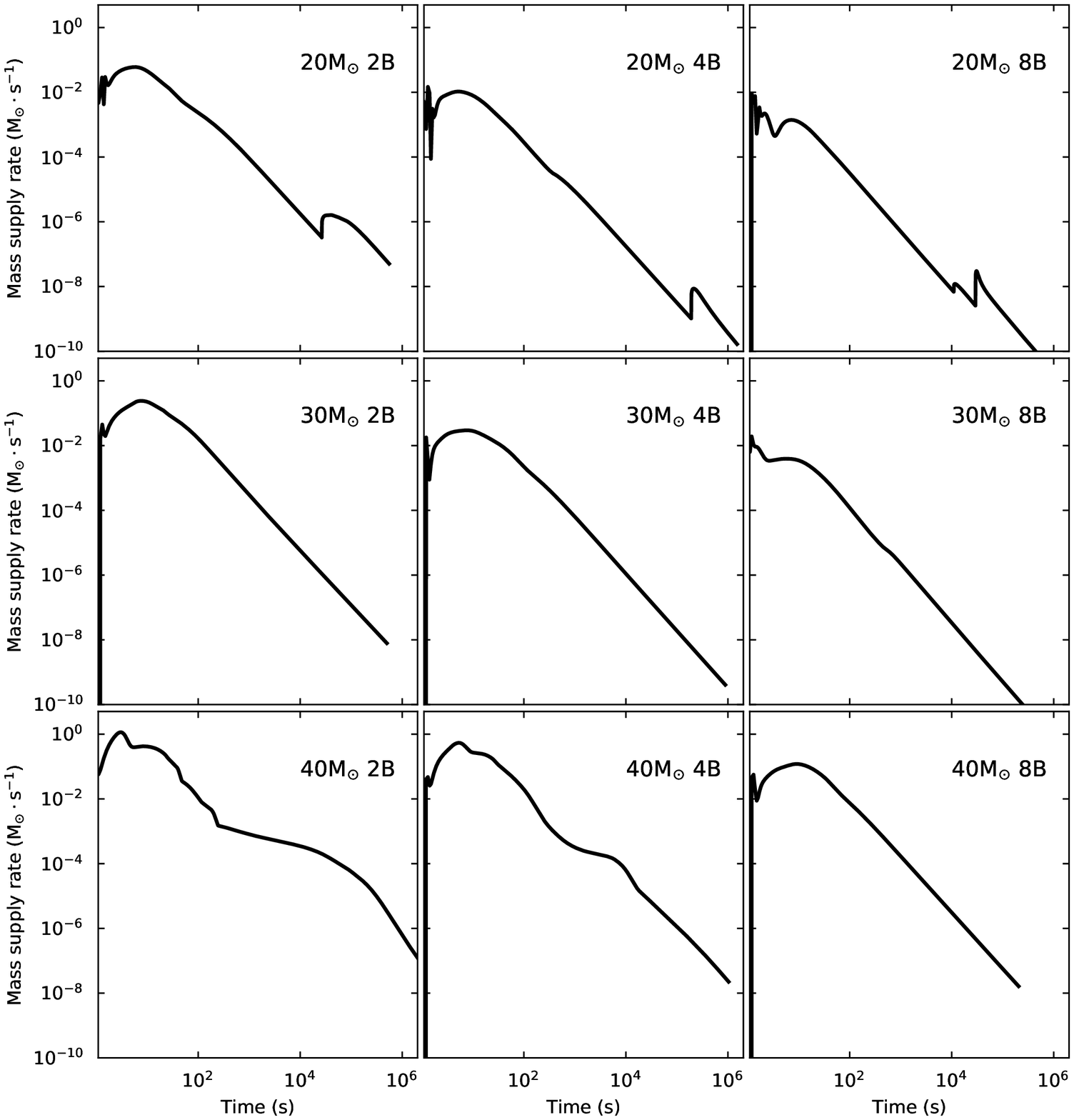}
 \caption{Time evolutions of mass supply rates with the progenitor metallicity $Z/Z_\odot$ = 0.01.}
\end{figure*}

\begin{figure*}
 \ContinuedFloat
 \includegraphics[width=10cm,height=10cm]{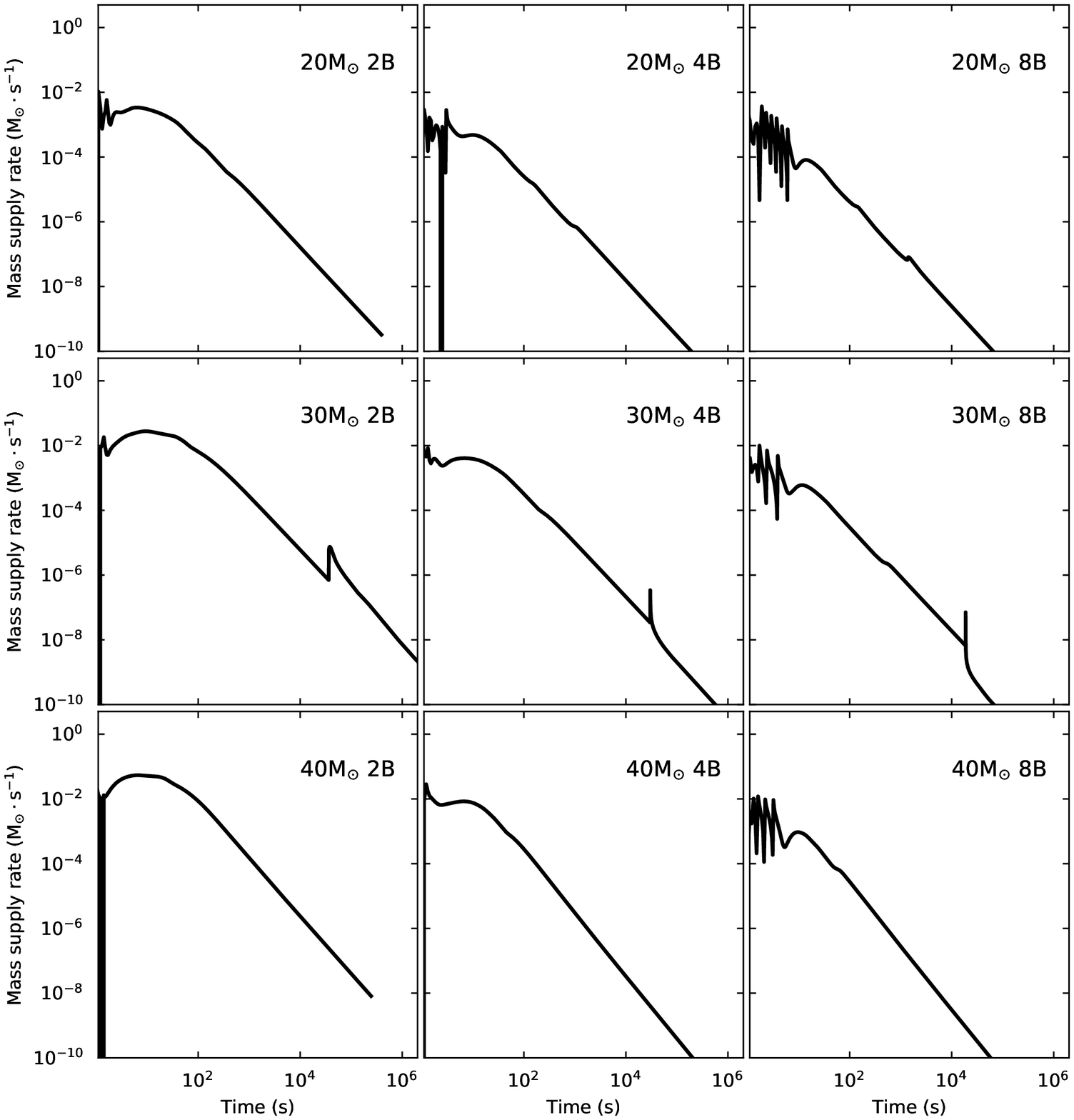}
 \caption{Time evolutions of mass supply rates with the progenitor metallicity $Z/Z_\odot$ = 1.}
\end{figure*}

\end{document}